\documentstyle[12pt]{article}
\renewcommand{\theequation}{\arabic{section}.\arabic{equation}}
\begin{document}

\topmargin=-2.0cm
\evensidemargin=0cm
\oddsidemargin=0cm

\date{}

\baselineskip=0.8cm

\renewcommand{\appendix}{\renewcommand{\thesection}
{Appendix}\renewcommand{\theequation}{\Alph{section}.\arabic{equation}}
\setcounter{equation}{0}\setcounter{section}{0}}

\vspace*{-2cm}
\begin{flushright}UT-Komaba 96-15
\end{flushright}

\vspace{2cm}

\begin{center}
{\Large{\bf Dynamical Symmetry Breaking\\
 in Light-Front Gross-Neveu Model}}

\vspace{1cm}

{\large Kazunori Itakura}\footnote{Electronic address: itakura@hep1.c.u-tokyo.ac.jp}\\
\vspace{1cm}
{\it Institute of Physics, University of Tokyo,}\\
{\it Komaba, Meguro-ku, Tokyo 153, Japan}
\vspace{2.5cm}

{\bf Abstract}
\end{center}
We investigate dynamical symmetry breaking 
 of the Gross-Neveu model in the light-front formalism 
 without introducing auxiliary fields.
While this system cannot have zero-mode constraints, we find that
a nontrivial solution to the constraint on nondynamical spinor fields
is responsible for symmetry breaking. 
The fermionic constraint is solved by systematic 1/$N$ expansion
using the boson expansion method as a technique.
Carefully treating the infrared divergence,  
we obtain a nonzero vacuum expectation value 
for fermion condensate in the leading order.
We derive the 't Hooft equation including the effect of condensation, 
and determine  the Hamiltonian consistently with the equation of motion.
 
\newpage

\section{Introduction}
One of the most crucial problems in the light-front (LF) formalism is 
how to describe spontaneously symmetry breaking on the {\it trivial} vacuum.
The idea that  solving a constraint on the longitudinal 
zero mode (zero-mode constraint) may provide us with 
mechanism for it  \cite{MasYama}
has been examined by several authors in 1+1 dimensional 
scalar models \cite{Pinsky,Heinzl}.
If the zero mode has a nontrivial $c$-number part, 
it gives a nonzero vacuum expectation value of fields. 
Although the nonperturbative calculation  is still difficult in 3+1 dimensions,
this approach is indispensable for  
 comparison with the other approach \cite{Wilson,Perry_Wilson} 
where , instead of solving zero mode constraints, 
 nontrivial vacuum effects are supposed to be able to be incorporated by 
suitable counter terms with symmetry consideration.
Then, from the standpoint of the zero-mode approach, 
 a  straightforward question arises: 
if there exists no bosonic field in the system,
how one can describe the symmetry breaking 
even without zero-mode constraints.
This question motivated us to investigate the simplest fermionic model, 
the (massive) Gross-Neveu model with global U($N$) symmetry \cite{Gross_Neveu}:
\begin{equation}
{\cal L}=\bar\Psi(i\gamma^{\mu}\partial_\mu-m_0)\Psi +
\frac{g^2}{2}(\bar\Psi\Psi)^2.
\end{equation}
In the equal-time formulation,  
we know that when the bare mass is absent, 
the discrete symmetry  $\Psi\rightarrow \gamma^5 \Psi$ breaks
 dynamically  via $\langle \bar\Psi \Psi\rangle \neq 0$.
So far the study of the light-front Gross-Neveu model
has been a little controversial.
First, the same phenomenon was successfully observed 
using the light-like quantization surface 
which approaches the light-cone surface as a limit  \cite{Hornbostel,Sakai}.
On the other hand, exactly on the light-front, 
it has been an unsettled problem whether we can obtain a nonzero condensate or not. 
as well as  the renormalization procedure \cite{Thies_Ohta,Pesando}.
However, Maedan \cite{Maedan1,Maedan2} obtained 
the same result as that of the equal-time formulation.
He solved the zero mode constraint, 
which was enabled by the introduction of a bosonic auxiliary field.

In this letter, 
we shall again make an analysis of the Gross-Neveu model 
without auxiliary fields included.
To compare with the previous works and to see carefully the way 
 condensation emerges, we put the system in a box of 
length $L$ and impose antiperiodic boundary condition.

\section{Formulation of the problem}
\setcounter{equation}{0}

The Gross-Neveu model on the LF has two features 
characteristic of the four-fermi interaction.
In the LF formalism,  half of the  spinor field is 
a dependent variable and should be constrained.
The constraint in the Gross-Neveu  model, however, 
 is difficult  to solve due to the nonlinearity of the four-fermi interaction.
Indeed, the Euler-Lagrange equations are given by
\begin{eqnarray}
i\partial_+ \psi &=&\left \{ \frac{m_0}{\sqrt{2}} 
- \frac{g^2}{2}(\psi^{\dagger}\chi+\chi^{\dagger}\psi)\right \}\chi, 
\label{eqofmotion}\\
i\partial_- \chi&=&\left \{ \frac{m_0}{\sqrt{2}} 
- \frac{g^2}{2}(\psi^{\dagger}\chi+\chi^{\dagger}\psi)\right \}\psi,
\label{constraint}
\end{eqnarray}
where 
$\Psi_a=2^{-1/4}(   \psi_a , \chi_a  )^{T},\ \ (a=1,\ldots,N)$
and  the light-front derivatives are $\partial_{\pm}=\partial/\partial x^{\pm}, 
x^{\pm}=(x^0\pm x^1)/\sqrt{2},\ \   x^-\in [-L,L],$
and the $\gamma$ matrices 
$\gamma_0=\sigma_1,\gamma_1=i\sigma_2.$
Equation (\ref{constraint}) includes only 
spatial derivative and thus is a constraint,
which  we call hereafter as a "fermionic constraint".

The Gross-Neveu model can also be defined as a limit of the 
Yukawa-like theory \cite{Maedan2}.
At the limit, the scalar field of the system becomes an auxiliary field.
There exist two constraints.
One is the zero-mode constraint for the auxiliary field 
and the other is  the fermionic constraint 
which is linear in terms of the spinor fields.
While the fermionic constraint is easily solved,
the zero-mode constraint is complicated and 
gives a nonzero value of condensation.
However, the solution to the fermionic constraint 
involves the zero-mode of the 
auxiliary field and, thus in a strict sense,  it is not solved at the first step.
What was done  in \cite{Maedan2} 
was to eliminate the nondynamical degrees of freedom from the 
coupled equations  by  a two-step approach.
Therefore we expect that, {\it without auxiliary fields, 
the fermionic constraint must be responsible for dynamical symmetry breaking} 
and so we shall carefully solve eq. (\ref{constraint}).

Another problem
is vanishing of the classical Hamiltonian in the massless limit.
According to  Dirac's procedure, 
the  Hamiltonian  on the constrained surface is given by 
\begin{equation}
H=\frac{m_0}{2\sqrt{2}}\int_{-L}^L dx^- 
(\psi^{\dagger}\chi+ \chi^{\dagger}\psi),
\end{equation}
where $\chi$ should be substituted by a solution of (\ref{constraint}).
Thus, the Hamiltonian vanishes in the massless limit \cite{Pesando}. 
However, if the spontaneous symmetry breaking occurs, 
the fermion acquires a nonzero mass
and even if  the bare mass goes to zero, 
the massless limit must have nontrivial Hamiltonian.
Nevertheless, it would be difficult to observe 
such effects in the classical Hamiltonian 
since the symmetry breaking in the Gross-Neveu model
takes place  only in the quantum level.
Therefore we decide to start from the Euler-Lagrange 
 equations as in Ref. \cite{Thies_Ohta}. 
The Hamiltonian will be constructed consistently 
with the equation of motion 
after we solve the fermionic constraint. 

Since the fermion condensate is given as the vacuum expectation value
 of the fermion bilinear operator,
it would be better to rewrite (\ref{eqofmotion}) 
and (\ref{constraint}) in terms of   bifermion operators.
We introduce U($N$) singlet bilocal operators at the equal light-front time:
\begin{eqnarray}
M(x^-,y^-;x^+)&=&\sum_{a=1}^N \psi^{\dagger}_a(x^-,x^+)\psi_a(y^-,x^+),\\
C(x^-,y^-;x^+)&=&\sum_{a=1}^N \psi^{\dagger}_a(x^-,x^+)\chi_a(y^-,x^+).
\end{eqnarray}
We also introduce $T(x,y)=C(x,y)+C^{\dagger}(x,y)$ for convenience. From 
now on, we work at the fixed light-front time and 
omit $x^+$-dependence and superscripts for the light-front spatial variables
unless needed.

The equations for the bilocal operators are written as
\begin{equation}
i\partial_+M(x,y)=\frac{m_0}{\sqrt{2}}
\Big( C(x,y)- C^{\dagger}(y,x)\Big) -\frac{g^2}{2}\Big (C(x,y)T(y,y)-T(x,x)C^{\dagger}(y,x)\Big),\label{eq_of_motion} \\
\end{equation}
\begin{equation}
i\frac{\partial}{\partial y^-}T(x,y)=\frac{m_0}{\sqrt{2}}
\Big(M(x,y)-M(y,x)\Big)-\frac{g^2}{2}
\Big(M(x,y)T(y,y)-T(y,y)M(y,x)\Big)
. \label{constraint2}
\end{equation}
We define the theory by these equations 
with this ordering and a quantization condition
on the dynamical fermion:
\begin{equation}
\{\psi_a(x),\psi^{\dagger}_b(y)\}_{x^+=y^+}=\delta_{ab}\delta(x^--y^-),\ \ 
\{\psi_a(x), \psi_b(y)\}_{x^+=y^+}=0.
\end{equation}
Fourier expansions of the bifermion operators are defined by
\begin{equation}
M(x,y)=\frac{1}{2L}\sum_{n,m\in {\bf Z}+\frac{1}{2}}
e^{-ik_n x}e^{-ik_m y}M(n,m),
\end{equation}
and so on, where  $k^+_n=\pi n/L$ and $n, m$ are half integers 
due to  the anti-periodic boundary condition
$\psi(L)=-\psi(-L)$.
Momentum representation of the above equations are 
\begin{eqnarray}
i\partial_+M(n,m)&=&\frac{m_0}{\sqrt{2}}
\Big(C(n,m)-C^{\dagger}(-m,-n)\Big)\nonumber\\
&-&\frac{g^2}{4L}\sum_{k,l} \left( C(n,m-k-l)T(k,l)
-T(k,l)C^{\dagger}(-m,k+l-n)\right), \label{eqofmo_mom}
\end{eqnarray}
and 
\begin{eqnarray}
k_m T (n,m)&=&\frac{m_0}{\sqrt{2}}\Big(M(n,m)-M(m,n) \Big)\nonumber\\
&-&\frac{g^2}{4L}\sum_{k,l\in {\bf Z}+\frac{1}{2}}\Big(M(n,m-k-l)T(k,l) 
-T(k,l)M(m-k-l,n)\Big).\label{const_mom}
\end{eqnarray}

\section{Solution to the fermion constraint}
\setcounter{equation}{0}
We solve the fermionic constraint (\ref{const_mom})  
by using the 1/$N$ expansion.
Let us expand the bilocal operators as
\begin{equation}
M(n,m)=N\sum_{p=0}^{\infty}\left(\frac{1}{\sqrt{N}}\right)^p
\mu^{(p)}(n,m),
\end{equation}
and  similarly, $C(n,m)$ and $T(n,m)$ are expanded by 
$c^{(p)}(n,m)$ and $t^{(p)}(n,m)$, respectively. 
Expansion of $M(n,m)$ can be given by the boson expansion method 
which is a familiar technique in the many-body physics \cite{itakura}.
Among various ways of the boson expansions, 
the Holstein-Primakoff type  is the most useful  
for large $N$ theories.
Introducing bosonic operators
\begin{equation}
\Big[B(n_1,n_2), \ B^{\dagger}(m_1,m_2)\Big]=
\delta_{n_1,m_1}\delta_{n_2,m_2},\ \ 
\Big[B(n_1,n_2),\  B(m_1,m_2)\Big]=0,
\end{equation}
$M(n,m)$'s are represented as follows:
\begin{eqnarray}
:M_{-+}(n_1,n_2):&=&\sum_{k=\frac{1}{2},\frac{3}{2},\cdots}
B^{\dagger}(-n_1,k)B(n_2,k)\equiv {\cal A}(n_2,-n_1),\\
:M_{+-}(n_1,n_2):&=&-\sum_{k=\frac{1}{2},\frac{3}{2},\cdots}
B^{\dagger}(k,-n_2)B(k,n_1),\\
:M_{++}(n_1,n_2):&=&\sum_{k=\frac{1}{2},\frac{3}{2},\cdots}
(\sqrt{N-{\cal A}})(n_2,k)B(k,n_1),\label{BEM3}\\
:M_{--}(n_1,n_2):&=&\sum_{k=\frac{1}{2},\frac{3}{2},\cdots}
B^{\dagger}(k,-n_2)(\sqrt{N-{\cal A}})(k,-n_1),\label{BEM4}
\end{eqnarray}
where the suffices imply the sign of the momentum and
the normal order is defined on the Fock vacuum.
The right-hand-sides of these are determined 
so that this representation  satisfies the algebra of $:M(n,m):$.
If we expand eqs. (\ref{BEM3}) and (\ref{BEM4}) in terms of 1/$N$,
each order is explicitly given by $B$ and $B^{\dagger}$ 
unlike the naive 1/$N$ expansion of $M(n,m)$ in Ref. \cite{Thies_Ohta}. 
Therefore, we can express $c^{(n)}$ also 
 in terms of the bosonic operators $B$ and $B^{\dagger}$ in principle.

Using $\mu^{(0)}(n,m)=\theta(n) \delta_{n+m, 0}$, 
the lowest order of the fermion constraint is given by
\begin{equation}
k_mt^{(0)}(n,m)=\frac{m_0}{\sqrt{2}}\epsilon(n)
\delta_{n+m, 0}
-\frac{g^2_0}{4L}\epsilon(n)\sum_{k\in{\bf Z}+\frac{1}{2}}
t^{(0)}(k,n+m-k),\label{lowest_constraint}
\end{equation}
where $g_0^2=g^2N$ and $\epsilon(n)=\theta(n)-\theta(-n)$.
We find that $t^{(0)}(n,m)$ is a $c$-number because there are
 no operators in this equation. 
The $c$-number part of a bilocal operator is a function of only $x-y$ 
due to the translational invariance of the vacuum, which implies
$t^{(0)}(n,m)=t^{(0)}(n,-n)\delta_{n+m,0}$.

By the way, if the lowest order equation has a nontrivial solution,
it gives rise to the physical fermion mass 
\begin{equation}
M\equiv m_0-g^2\langle \bar\Psi\Psi\rangle =
 m_0-\frac{g^2_0}{2\sqrt{2}L}\sum_{k\in{\bf Z}+\frac{1}{2}}
t^{(0)}(k,-k).\label{phys_mass}
\end{equation}
Inserting this and $m=-n$ into (\ref{lowest_constraint}),  
and summing over $n\in {\bf Z}+1/2$, 
we obtain an equation
\begin{equation}
\frac{1}{g_0^2}-\frac{1}{2\pi}\frac{M}{M-m_0}
\sum_{n=\frac12, \frac32,\cdots}\frac{\Delta k}{k_n}=0,
\label{lowest_constraint_2}
\end{equation}
where  $\Delta k=\pi/L$.
If we set $m_0=0$ naively, this equation seems to become independent of $M$
and thus we cannot determine $M$ by this equation.
Or if we solve (\ref{lowest_constraint_2}) in terms of $M$, 
and take the $m_0\rightarrow 0$ limit, $M$ also seems to vanish. 
However,  this observation is not correct 
because this equation is not well-defined until the summation is regularized.
Indeed, by carefully treating the divergences, this equation gives a  
nonzero value for $M$ even in the $m_0=0$ case.

Since eq. (\ref{lowest_constraint_2}) is independent of the box length $L$,
we can evaluate it in any box size.
Nevertheless we evaluate it in the continuum limit $L\rightarrow \infty$ 
because it is what we want eventually.
Then, we find that the summation has both infrared and ultraviolet divergences.
We regularize it as follows:
\begin{equation}
\sum_{n=\frac12,\frac32,\cdots}\frac{\Delta k}{k_n}=
\sum_{m\ge 0 }\frac{1}{m+1/2}\ \longrightarrow \ 
\int_{\frac{M^2}{2\Lambda}}^{\Lambda} \frac{dk}{k}=
{\rm ln} \frac{2\Lambda^2}{M^2}.
\end{equation}
This is considered to be the parity invariant regularization.
Since the parity transformation is the exchange of $k_-$ and $k_+$,
the cut off $k_-<\Lambda$ should be paired with $k_+=M^2/2k_-<\Lambda$,
which inevitably relates the UV and IR cutoffs.
Here the dispersion of the fermion is thought to be $2k_+k_-=M^2$
because the fermion acquires the physical mass (\ref{phys_mass}).
This procedure corresponds to imposing a consistency condition.
If we introduce  UV and IR cutoffs independently, 
it generally breaks the parity invariance and we cannot yield the correct result.
It should be commented that eq. (\ref{lowest_constraint_2}) in the massless 
case is identical with the leading order of the zero mode constraint 
in the auxiliary field approach \cite{Maedan2}.
There, the infrared regularization was performed by the heat kernel method, 
which gives the same result. 

Renormalization is implemented as follows.
The coupling constant is renormalized  as 
\begin{equation}
\frac{1}{g_R^2(\mu)}=\frac{1}{g_0^2}
-\frac{1}{2\pi}{\rm ln}\frac{2\Lambda^2}{\mu^2}+\frac{1}{\pi},
\end{equation}
\begin{equation}
g_0^2=Z(\mu)g_R^2(\mu), \ \ \ 
 Z^{-1}(\mu)=1+g_R^2\left( \frac{1}{2\pi}{\rm ln} 
\frac{2\Lambda^2}{\mu^2}-\frac{1}{\pi}\right),
\end{equation}
which shows the asymptotic freedom.
Mass  can be renormalized by the same  factor: 
$ m_0=Z(\mu) m_R(\mu)$.
Rewriting eq. (\ref{lowest_constraint_2}) in terms of the renormalized 
quantities, we obtain an equation for the physical fermion mass;
\begin{equation}
{\rm ln} \frac{M^2}{\mu^2}=2-\frac{M-m_R}{M}\frac{2\pi}{g_R^2}.\label{final}
\end{equation}
In the $m_R=0$ case,
the solution is 
\begin{equation}
M=\mu \exp\left\{1-\frac{\pi}{g_R^2}\right\}\equiv M_0.\label{massless}
\end{equation}
This is the same result as that of the original work by 
Gross and Neveu \cite{Gross_Neveu}. In the massive case,
using the renormalization invariant  parameter $M_0$,  
eq. (\ref{final}) can be written as
\begin{equation}
{\rm ln}\frac{M^2}{M_0^2}=\frac{2\pi}{g_R^2}\frac{m_R}{M}.
\end{equation}
This equation has a nonvanishing solution. 
If $m_R/M_0\ll 1$, it is $M\simeq M_0+(\pi/g_R^2)m_R$,
which smoothly goes to $M_0$ as $m_R\rightarrow 0$.
Eventually,  $c^{(0)}$ is  determined  from eq. (\ref{lowest_constraint}) as
\begin{equation}
c^{(0)}(n,m)=-\frac{M}{\sqrt{2}}\frac{\theta(n)}{k_n}\delta_{n+m,0}.
\end{equation}

Similarly, the  fermionic constraint in the next leading order 
\begin{eqnarray}
k_mt^{(1)}(n,m)&=&\frac{M}{\sqrt{2}}
\Big(\mu^{(1)}(n,m)-\mu^{(1)}(m,n)\Big)\nonumber\\
&-&\frac{g_0^2}{4L}\epsilon(n)
\sum_{k\in {\bf Z}+\frac{1}{2}}t^{(1)}(k,n+m-k), 
\end{eqnarray}
is solved as
\begin{eqnarray}
c^{(1)}(n,m)&=&\frac{M}{\sqrt{2}}\frac{1}{k_m}  \mu^{(1)}(n,m) \\
&+&\left(  \frac{m_R}{M}+
\frac{g_R^2}{2L}\sum_{0<n<|K|}\frac{1}{k_n}  \right)^{-1} 
\frac{1}{4L}\frac{M}{\sqrt{2}}\frac{\theta(n)}{k_m}
\sum_{k\in {\bf Z}+\frac{1}{2}}
\frac{1}{k_{k-K}}\Big(\mu^{(1)}(k,K-k)-\mu^{(1)}(K-k,k)\Big) , \nonumber
\end{eqnarray}
where $K=n+m$.
This can be expressed in terms of $B(n,m)$ and $B^{\dagger}(n,m)$
by  using 
\begin{eqnarray}
\mu^{(1)}_{++}(n,m)=B(m,n),&&
\mu^{(1)}_{--}(n,m)=B^{\dagger}(-n,-m), \label{nextlowest}\\
\mu^{(1)}_{+-}(n,m)=0,&&\mu^{(1)}_{-+}(n,m)=0.
\end{eqnarray}
Note that there is no mass correction in this order due to $\mu^{(1)}(n,-n)=0$.
Also  we can  easily obtain the higher orders  because the equation is always 
linear in the  highest order $c^{(p)}$. 
Thus $C(n,m)$ can be represented only by the bosonic variables 
$B(n,m)$ and $B^{\dagger}(n,m)$.

\section{Hamiltonian and the 't Hooft Equation}
\setcounter{equation}{0}

We explicitly solved the fermionic constraint up to the next leading order 
and $C(n,m)$ can be represented by $B$ and $B^{\dagger}$.
The next work is to rewrite the equation of motion 
in terms of the dynamical variables
and construct the Hamiltonian from it.

The nontrivial leading contribution of the equation of 
motion (\ref{eqofmo_mom}) is
\begin{eqnarray}
i\partial_+\mu^{(1)}(n,m)&=&
\frac{M^2}{2}\left(\frac{1}{k_n}+\frac{1}{k_m}\right)
\mu^{(1)}(n,m)\nonumber\\
&+&\frac{g_R^2}{4L}\frac{M}{\sqrt{2}}\Big(\theta(n)-\theta(-m)\Big)
\left(\frac{1}{k_n}-\frac{1}{k_m}\right){\cal F}(n+m),
\end{eqnarray}
where 
\begin{eqnarray}
{\cal F}(K)&=&\left(  \frac{m_R}{M}+\frac{g_R^2}{2L}
\sum_{0<n<|K|}\frac{1}{k_n}  \right)^{-1}\nonumber \\
&&\times \frac{M}{\sqrt{2}}\sum_{n\in {\bf Z}+\frac{1}{2}}
\frac{1}{k_{K-n}}\Big(\mu^{(1)}(n,K-n)-\mu^{(1)}(K-n,n)\Big).
\end{eqnarray}
Rewriting this in terms of the bosonic variables, 
we obtain equations of motion for $B(n,m)$ and $B^{\dagger}(n,m)$:
\begin{eqnarray}
i\partial_+B(n,m)&=&\sum_{k,l,n,m>0}K^{n,m}_{k,l}B(k,l),\label{eq1}\\
i\partial_+B^{\dagger}(n,m)&=&
-\sum_{k,l,n,m>0}K^{n,m}_{k,l}B^{\dagger}(k,l),\label{eq2}
\end{eqnarray}
where the  matrix is
\begin{eqnarray}
K_{k,l}^{n,m}&=&\frac{M^2}{2}\left(\frac{1}{k_n}
+\frac{1}{k_m}\right)\delta_{n,k}\delta_{m,l}\nonumber \\
&-&\left( \frac{m_R}{M}+\frac{g_R^2}{2\pi}\sum_{0<m<|K|}
\frac{\Delta k}{k_m}\right)^{-1}\frac{g_R^2}{4L}\frac{M^2}{2}
\left(\frac{1}{k_n}-\frac{1}{k_m}\right)\left(\frac{1}{k_k}
-\frac{1}{k_l}\right)\delta_{k+l,n+m}\label{kernel}.
\end{eqnarray}
It is easy to find the Hamiltonian which gives  these as the Heisenberg
 equations of motions: 
\begin{equation}
H=\sum_{n,m,k,l>0}B^{\dagger}(n,m)K^{n,m}_{k,l}B(k,l) + O(N^{-1/2}).
\end{equation}

Let us introduce the collective operators 
\begin{equation}
b^{(\alpha)\dagger}_K=\sum_{0<n<K}\Phi^{(\alpha)}(n)B^{\dagger}(n,K-n), \ \ \
 (\alpha=1,\ldots,K),
\end{equation}
where the wave function $\Phi^{(\alpha)}(n)$ is assumed to satisfy
the orthogonality and the  completeness.
Then $b_K$ satisfy the bosonic relations
$[b_K^{(\alpha)},b_{K'}^{(\alpha')\dagger}]=\delta_{\alpha, \alpha'}
\delta_{K,K'}$.
Note that $b_K$ does not have the zero mode because both of 
the arguments of $B$ should be positive.
The wave functions are determined so as to diagonalize the Hamiltonian.
The  eigenvalue equation  gives the 't Hooft equation.
If we introduce rescaled variables
$x_n\equiv k_n/k_K$ for $0<n<K$,
we obtain  
\begin{eqnarray}
2P_+k_K\Phi(n)&=&M^2\left(\frac{1}{x_n}+\frac{1}{1-x_n}\right)\Phi(n) \\
&-&\left( \frac{m_R}{M}+\frac{g_R^2}{2\pi}\sum_{0<m<K}\frac{\Delta x}{x_m}\right)^{-1}\frac{g_R^2M^2}{4\pi}\left(\frac{1}{x_n}-
\frac{1}{1-x_n}\right)\sum_{0<l<K}\Delta x\left(\frac{1}{x_l}-
\frac{1}{1-x_l}\right)\Phi(l),\nonumber
\end{eqnarray}
where we take the time dependence of $B^{\dagger}$ to be $e^{iP_+x^+}$.
Taking the continuum limit 
$\lim_{K,L\rightarrow \infty}\pi K/L=P_-,$
the 't Hooft equation becomes
\begin{eqnarray}
2P_+P_-\phi(x)&=&M^2\left(\frac{1}{x}+\frac{1}{1-x}\right)\phi(x) \label{tHooft}\\
&-&\left( \frac{m_R}{M}+\frac{g_R^2}{2\pi}\int_0^1\frac{dx}{x}\right)^{-1}
\frac{g_R^2M^2}{4\pi}\left(\frac{1}{x}-\frac{1}{1-x}\right)\int_0^1 dy\left(\frac{1}{y}-\frac{1}{1-y}\right)\phi(y).\nonumber
\end{eqnarray}
One can easily observe that when $x=1/2$, 
the second term of the 't Hooft equation vanish.
Thus the invariant mass of this state is
$2M$ which is known to exist as 
Gross and Neveu discussed \cite{Gross_Neveu}.
The wave function of this state is 
$\phi(x)=\delta(x-\frac{1}{2})$ and far from the collective one.
In this order, the state $b_K^{\dagger}|0\rangle=B^{\dagger}(K/2,K/2)|0\rangle$
can be easily translated into a bifermion state by using  eq. (\ref{nextlowest}), 
which means that the state can be understood as the constituent state.

The result (\ref{tHooft}) is different from that of \cite{Thies_Ohta} 
by the divergent factor of the second term, which, however, 
does not affect the above solution because of the vanishing of the second term.
This discrepancy originates from the different renormalization procedures.
If we renormalize only the UV divergence, 
we do not have such a divergent factor \cite{Thies_Ohta}.
The divergent factor seems to be inevitable in our renormalization prescription. 
However we expect that it cancels with the possible 
infrared divergence of the last integral.
Thus our treatment is different from the usual way of the 't Hooft equation.
Our expectation may be strengthened by a little more nontrivial example
in the chiral Gross-Neveu model 
${\cal L}_{\rm int}=g^2/2[(\bar\Psi\Psi)^2-(\bar\Psi\gamma^5\Psi)^2]$.
There is the chiral symmetry when $m_0=0$ and 
if it breaks spontaneously, there appears a Nambu-Goldstone boson.
Although there is no NG boson in two-dimensional field theories 
\cite{Coleman,Root},
we can observe it in the leading order.
In the same way as that of the Gross-Neveu model, 
the 't Hooft equation becomes
\begin{eqnarray}
2P_+P_-\phi(x)&=&M^2\left(\frac{1}{x}+\frac{1}{1-x}\right)\phi(x) \\
&-&\left( \frac{m_R}{M}+\frac{g_R^2}{2\pi}
\int_0^1\frac{dx}{x}\right)^{-1}\frac{g_R^2M^2}{2\pi}\int_0^1 dy\left(\frac{1}{xy}+\frac{1}{(1-x)(1-y)}\right)\phi(y).\nonumber
\end{eqnarray}
Although there emerges in the second term the same factor as 
in eq. (\ref{tHooft}), we find that $\phi(x)=constant$ is the exact solution
for massless case.
The invariant mass of this state is zero and this solution corresponds to 
the Nambu-Goldstone boson.
Precisely the divergence of the last integral cancels with the divergent factor.

\section{Discussion}
We calculated the nonzero value of  condensation in the Gross-Neveu model
on the light front without introducing auxiliary fields.
We found that the nontrivial solution to the fermionic constraint 
lead to the fermion condensation.
In solving the fermionic constraint we used the boson 
expansion method for the systematic 1/$N$ expansion, 
although it was not essential for the analysis up to the 
next leading order.

To obtain the nonzero value of condensation, 
it is very important to treat  the infrared divergence with great care. 
It seems suggestive that the prescription for the infrared divergence
can be set to be related with symmetry such as  the parity invariance. 
Such symmetry consideration will 
be indispensable to obtain the correct value of the condensation, 
which will be true for another approach without zero-modes 
\cite{Wilson,Perry_Wilson}. 

The emergence of the nonzero condensate
 is connected with the renormalization prescription
and affects  on  the 't Hooft equation.  
That is, the equation cannot help having a divergent factor.
However, the 't Hooft equations we obtained has  
not only a well known solution  in the Gross-Neveu model  but also 
a solution corresponding to the Nambu-Goldstone mode 
in the chiral Gross-Neveu model.
Usually,  the 't Hooft equation is derived without effects of condensation
included and the fermion condensate is obtained by, for example, 
the sum-rule calculation \cite{Zhit_Br}.
It would be interesting to investigate the relation between two approaches.

We constructed the Hamiltonian consistently with the equation of motion 
after solving the fermionic constraints.
Therefor we cannot calculate the vacuum energy and cannot determine 
which vacuum is realized. 
Indeed, there is also a symmetric solution, $M=0$ in the $m_R=0$ limit.
However, in this case, if we start from the very massive fermion and decrease the mass, 
 we can uniquely determine the vacuum and it is the broken phase.\\

{\bf Acknowledgments}\\
The author would like to thank K. Harada and S. Tsujimaru 
for stimulating discussions. 
Special thanks are due to S. Maedan 
for valuable comments and suggestions.
And he also acknowledges the kind hospitality 
of the particle theory group in Kyushu University.

\end{document}